\documentclass[sigconf, dvipsnames, language=Chinese, language=English]{acmart}

\AtBeginDocument{%
  \providecommand\BibTeX{{%
    \normalfont B\kern-0.5em{\scshape i\kern-0.25em b}\kern-0.8em\TeX}}}

\setcopyright{acmcopyright}

\acmConference[ESEC/FSE 2022]{The 30th ACM Joint European Software Engineering Conference and Symposium on the Foundations of Software Engineering}{14 - 18 November, 2022}{Singapore}

\usepackage{CJKutf8}
\usepackage{booktabs}
\usepackage{multicol}
\usepackage{multirow}
\usepackage{makecell}
\usepackage{caption}
\usepackage{subcaption}
\usepackage{xcolor}
\usepackage{pgfplots}
\usepackage{bbding}

\newcommand{\st}{\vspace{-1.8em}}

\begin{document}


\title{Incorporating Domain Knowledge through Task Augmentation for Front-End JavaScript Code Generation}


\author{Sijie Shen}
\affiliation{%
    \institution{Key Lab of High Confidence Software Technology, MoE (Peking University)}
    \city{Beijing}
    \country{China}}
\email{sjshen@pku.edu.cn}

\author{Xiang Zhu}
\affiliation{%
    \institution{Alibaba Group}
    \city{Hangzhou}
    \country{China}}
\email{hanling.zx@alibaba-inc.com}

\author{Yihong Dong}
\affiliation{%
    \institution{Key Lab of High Confidence Software Technology, MoE (Peking University)}
    \city{Beijing}
    \country{China}}
\email{dongyh@stu.pku.edu.cn}

\author{Qizhi Guo}
\affiliation{%
    \institution{Alibaba Group}
    \city{Hangzhou}
    \country{China}}
\email{qizhi.gqz@alibaba-inc.com}

\author{Yankun Zhen}
\affiliation{%
    \institution{Alibaba Group}
    \city{Hangzhou}
    \country{China}}
\email{zhenyankun.zyk@alibaba-inc.com}

\author{Ge Li}
\authornote{Corresponding author.}
\affiliation{%
    \institution{Key Lab of High Confidence Software Technology, MoE (Peking University)}
    \city{Beijing}
    \country{China}}
\email{lige@pku.edu.cn}

\renewcommand{\shortauthors}{Shen, et al.}


\begin{abstract}
    Code generation aims to generate a code snippet automatically from natural language descriptions. Generally, the mainstream code generation methods rely on a large amount of paired training data, including both the natural language description and the code. However, in some domain-specific scenarios, building such a large paired corpus for code generation is difficult because there is no directly available pairing data, and a lot of effort is required to manually write the code descriptions to construct a high-quality training dataset. Due to the limited training data, the generation model cannot be well trained and is likely to be overfitting, making the model's performance unsatisfactory for real-world use. To this end, in this paper, we propose a task augmentation method that incorporates domain knowledge into code generation models through auxiliary tasks and a Subtoken-TranX model by extending the original TranX model to support subtoken-level code generation. To verify our proposed approach, we collect a real-world code generation dataset
    and conduct experiments on it. Our experimental results demonstrate that the subtoken-level TranX model outperforms the original TranX model and the Transformer \cite{vaswani2017attention} model on our dataset, and the exact match accuracy of Subtoken-TranX improves significantly by 12.75\% with the help of our task augmentation method. The model performance on several code categories has satisfied the requirements for application in industrial systems. Our proposed approach has been adopted by Alibaba's \emph{BizCook} platform. To the best of our knowledge, this is the first domain code generation system adopted in industrial development environments.

\end{abstract}




\settopmatter{printfolios=true}
\maketitle

\section{Introduction}
\label{section:introduction}

Automatic code generation refers to generating code according to a function description in natural language. The code generation technology can improve the automation level of software development and reduce the workload of software developers, thereby effectively improving the efficiency and quality of software development and maintenance.
In recent years, researchers have proposed a series of deep-learning code generation methods with remarkable achievements, such as \cite{yin2018tranx,dong2018coarse,sun2019grammar,sun2020treegen}.

Despite the success of the above code generation methods, for all we know, we rarely see the application of code generation in existing industrial systems.
Generally, the mainstream code generation methods rely on the accessibility of abundant training data for their success.
However, the code in the industrial development environment is usually under a specific domain scenario, where directly paired training data is not always available.
In our scenario, we want to generate JavaScript expressions from natural language descriptions, and our major limitation is the insufficient training data problem.
In order to perform code generation, we have to build the training dataset from scratch, i.e., mine a group of code and manually write the corresponding semantic descriptions in natural language for each of them. The cost of this process is quite expensive for the construction of a high-quality dataset, so we can only obtain relatively small-scale paired training data. The lack of training data makes it hard for the neural networks to learn the effective representation to generate code, which dramatically decreases the model performance.

In this paper, we propose a task augmentation method to alleviate the negative effect of insufficient paired training data by incorporating domain knowledge into the code generation model.
We design some auxiliary tasks and utilize the domain knowledge that is easily acquired from the codebase and requirement documents as paired training data for the tasks.
By training the model together with the code generation main task and auxiliary tasks, the model is able to learn the domain knowledge from the extra tasks and improve its performance on code generation. Furthermore, we present the Subtoken-TranX model by extending the original TranX \cite{yin2018tranx} model to support subtoken-level code generation. The Subtoken-TranX model helps learn the token embedding well and captures the relationship of tokens containing the same subtoken from limited training data. We carry out a series of experiments to demonstrate the effectiveness of task augmentation with domain knowledge and the Subtoken-TranX model. After applying the task augmentation, the Subtoken-TranX achieves a top-1 exact match accuracy of 33.16\% and top-5 accuracy of 40.31\%, compared to 20.41\% and 29.08\%, respectively, without task augmentation. The Subtoken-TranX model also outperforms the original TranX model and Transformer model. The results on two categories of data have satisfied the requirements for application in actual industrial systems, and the \emph{BizCook} platform has adopted our approach.

Our contributions can be summarized below:
\begin{itemize}
    \item We propose a real-world code generation task and the corresponding dataset\footnote{Available online at \url{https://tianchi.aliyun.com/dataset/dataDetail?dataId=107819}.}. The dataset consists of 2,489 paired data of JavaScript expressions extracted from Alibaba's codebase and corresponding descriptions in the natural language of Chinese.
    \item We propose a task augmentation method to incorporate domain knowledge to code generation models through auxiliary tasks.
    \item We present the Subtoken-TranX model by extending the original TranX model to support subtoken-level code generation.
    \item We carry out experiments and demonstrate the effectiveness of the task augmentation method and Subtoken-TranX model. Our approach has been adopted by Alibaba's \emph{BizCook} platform and is continuously supporting front-end development.
\end{itemize}

\section{Background}
\label{section:background}


The application scenario of our work is the \emph{BizCook} platform in Alibaba. This platform is a front-end development platform tailored based on the business characteristics of \emph{Taobao}. This system covers the main stages of the entire process of front-end development, including requirements, design, coding, and testing. It aims to introduce intelligent approaches to the development process and improve development efficiency. Among them, an important direction of exploration is to generate code from requirement documents and design drafts.

In front-end development, most of the front-end code falls into these three categories:
\begin{itemize}
    \item User interface code. They are used to control the layout, style, etc., of UI elements, such as the position of a text area and the color of a button.
    \item Business logic code. They are used to control displayed content of UI elements, such as the text content in a text area or on a button.
    \item Control flow code. They are used to control the click or other behaviors of elements, such as the on-click event of a button.
\end{itemize}
Among the three categories, the user interface code is closely related to the design drafts. In our previous works, we have already developed a tool called \emph{imgcook}\footnote{\url{https://www.imgcook.com}.} to directly generate user interface code from Sketch, Photoshop, and Figma design drafts. On the basis of \emph{imgcook}, we want to extend the code generation ability of the \emph{BizCook} system to other categories. Since the descriptions of business logic code and control flow code are written in the requirement documents, this feature requires the ability to read the natural language descriptions in the requirement documents, convert them into JavaScript expressions, and bind the expressions to UI elements. It is a challenge to generate all types of front-end code simultaneously. Considering that the business logic code accounts for a large proportion of front-end code and is less complicated than the control flow code, we decide first to generate the business logic code at the present stage. This paper presents our practice in dealing with the task of generating the business logic JavaScript code from natural language descriptions.



\section{Dataset and Preprocessing}
\label{section:dataset}

\subsection{Code Generation Dataset}
\label{subsection:dataset:code_gen_data}

In the \emph{BizCook} platform, all the skeletons of code are predefined, and the platform provides ways to set the style, behavior, and display content. During coding, programmers only need to select a UI element and write the target style or string value expressions in JavaScript. The platform will insert the expressions into JavaScript code and bind them to the specific UI element. So all we need to generate is JavaScript expressions.
As we stated in the third paragraph of section \ref{section:background}, in the current stage, we only focus on the business logic code expressions that control the display text in the UI elements. Therefore, we randomly collect some expressions from Alibaba's codebase, filter and de-duplicate them, and get a collection of business logic code expressions. In order to get the complete paired data, we still have to manually produce the natural language description for each code. Finally, we obtain a paired dataset containing 2,489 examples.
We randomly split the dataset into a training set and a test set containing 2293 and 196 examples. When conducting experiments, we randomly sample a subset from the training set as the validation set. Table \ref{table:dataset_details} shows the details of the dataset.

\begin{table}
    \centering
    \begin{tabular}{lll}
        \toprule
                             & Train Set & Test Set \\
        \midrule
        \# Examples          & 2,293     & 196      \\
        \# Desc. Tokens      & 43,719    & 3,874    \\
        \# Code Tokens       & 25,806    & 2,256    \\
        \# Avg. Desc. Tokens & 19.07     & 19.77    \\
        \# Avg. Code Tokens  & 11.25     & 11.51    \\
        \bottomrule
    \end{tabular}
    \caption{Details of code generation dataset. Since the description is written in Chinese, the token numbers are counted as the number of Chinese characters. All details are counted after the preprocessing described in section \ref{subsection:dataset:preprocessing}.}
    \label{table:dataset_details}
    \st
\end{table}

In our dataset, we can further divide the JavaScript logic expressions into four categories:
\begin{itemize}
    \item string template expression (STE). Code of this category is string expressions produced by filling a string template with variables.
    \item OR logic expression (OLE). Code of this category are several values joined by short circuit OR operator.
    \item condition expression (CE). Code of this category is often a ternary conditional expression.
    \item data processing expression (DPE). Code of this category often contains processing to a data, such as taking a substring of a string.
\end{itemize}
Table \ref{table:data_example} shows examples of each category. In actual use, the input descriptions are in Chinese, and we provide the corresponding English translation here. We tag every piece of data in the test set with category labels and count the test set details by category. Table \ref{table:category_details} shows the details of test set by category.

\begin{CJK*}{UTF8}{gbsn}
\begin{table*}
    \centering
    \begin{tabular}{ll}
        \toprule
        Category     & String template expression                                                                                               \\
        Description  & 显示“优惠券已抵扣xx元”，xx为折扣价                                                                                       \\
        Desc. Trans. & Displays ``The coupon has been deducted xx yuan'', xx is the discounted price                                            \\
        Code         & \texttt{\{\textasciigrave 优惠券已抵扣\$\{discountPrice\}元\textasciigrave \}}                                           \\
        \midrule
        Category     & OR logic expression                                                                                                      \\
        Description  & 动态展示用户昵称，兜底为空                                                                                               \\
        Desc. Trans. & Dynamically display the user's nickname, the default value is empty                                                      \\
        Code         & \texttt{\{ user \&\& user.nick || " " \}}                                                                                \\
        \midrule
        Category     & Condition expression                                                                                                     \\
        Description  & 如果内容类型为直播，则展示直播时间描述，否则展示营销时间描述                                                             \\
        Desc. Trans. & If the content type is live, show the description of the live time, otherwise show the description of the marketing time \\
        Code         & \texttt{\{contentType === 'live' ? liveTimeDesc : marketingTimeDesc\}}                                                   \\
        \midrule
        Category     & Data processing expression                                                                                               \\
        Description  & 动态展示金币展示价格小数部分                                                                                             \\
        Desc. Trans. & Dynamically display the fractional part of coin show price                                                               \\
        Code         & \texttt{\{\textasciigrave\$\{data.coinShowPrice.split(".")[1]\}\textasciigrave\}}                                        \\
        \bottomrule
    \end{tabular}
    \caption{Examples of data in different categories.}
    \label{table:data_example}
\end{table*}
\end{CJK*}

\begin{table}
    \centering
    \begin{tabular}{lllll}
        \toprule
                             & ST    & OLE   & CE    & DPE   \\
        \midrule
        \# Examples          & 49    & 71    & 71    & 5     \\
        \# Desc. Tokens      & 662   & 1,242 & 1,856 & 114   \\
        \# Code Tokens       & 503   & 609   & 1,053 & 91    \\
        \# Avg. Desc. Tokens & 13.51 & 17.49 & 26.14 & 22.80 \\
        \# Avg. Code Tokens  & 10.27 & 8.58  & 14.83 & 18.20 \\
        \bottomrule
    \end{tabular}
    \caption{Details of test data by category.}
    \label{table:category_details}
    \st
\end{table}

\subsection{Preprocessing}
\label{subsection:dataset:preprocessing}
Considering that our dataset scale is small and the variability and noise in the dataset can affect code generation performance, we perform a series of preprocessing on the original data.
In this section, we introduce three main preprocessing methods we use.

\subsubsection{Code Canonicalization}

In our dataset, the code have some stylistic differences since the codes are not written by the same person. For example, developers have different preferences of using single and double quotes in string literals and whether to add semicolons at the end of statements, etc. We use the Esprima\footnote{\url{https://esprima.org}} parser to parse the JavaScript code into an abstract syntax tree (AST) and then use the Escodegen\footnote{\url{https://github.com/estools/escodegen}} code generator to convert the AST back to a canonicalized code. In this way, we can eliminate the stylistic differences between the codes to a certain extent.

\subsubsection{String Literal Replacement}

In our dataset, a proportion of code contains string literals. According to our regulations, all the string literals in the code must appear in the natural language description. Therefore, we can use placeholders to replace these string literals to simplify the code generation. As shown in the table \ref{table:simplify_const_str}, for string literals that do not contain variables, we use placeholders such as \texttt{<STR1>}, \texttt{<STR2>} to replace string literals. For string literals containing variables such as ``xx'', ``yy'', etc., we replace these string literals segment-wise with placeholders. These placeholders will be considered a single token by the model.

\begin{CJK*}{UTF8}{gbsn}
\begin{table*}
    \centering
    \begin{tabular}{p{0.18\textwidth}p{0.75\textwidth}}
        \toprule
        Original Description    & 判断是否幸运，条件成立则显示‘恭喜你押中啦’，否则显示‘很遗憾未押中’                                                                   \\
        Ori. Desc. Translation  & Test if lucky or not, display `Congratulations on your bet' if the conditions are met, otherwise display `Unfortunately not betting' \\
        Original Code           & \texttt{\{isLucky ? '\vphantom{}恭喜你押中啦' : '\vphantom{}很遗憾未押中'\}}                                                                               \\
        Simplified Description  & 判断是否幸运，条件成立则显示‘<STR1>’，否则显示‘<STR2>’                                                                               \\
        Simp. Desc. Translation & Test if lucky or not, display `<STR1>' if the conditions are met, otherwise display `<STR2>'                                         \\
        Simplified Code         & \texttt{\{isLucky ? '<STR1>' : '<STR2>'\}}                                                                                           \\
        \midrule
        Original Description    & 显示‘满xx使用’，xx为起步费                                                                                                           \\
        Ori. Desc. Translation  & Displays `For orders over xx, use the coupon', xx is the starting fee                                                                \\
        Original Code           & \texttt{\{'\vphantom{}满' + startFee + '\vphantom{}使用'\}}                                                                                                \\
        Simplified Description  & 显示‘<STR1> xx <STR2>’，xx为起步费                                                                                                   \\
        Simp. Desc. Translation & Displays `<STR1> xx <STR2>', xx is the starting fee                                                                                  \\
        Simplified Code         & \texttt{\{'<STR1>' + startFee + '<STR2>'\}}                                                                                          \\
        \bottomrule
    \end{tabular}
    \caption{Examples of string literal replacement}
    \label{table:simplify_const_str}
\end{table*}
\end{CJK*}

\subsubsection{Member Access Simplification}

A lot of code includes access to member variables. We simplified the member variable access in the code by removing the accessed object and only keeping the accessed fields. For examples, \texttt{task.status} is simplified to \texttt{status} and \texttt{task.assets.completeBtn} is simplified to \texttt{completeBtn}. We simplify the member access because we cannot predict the accessed object correctly from the description without programming context. Other modules in the development platform will search the programming context for the proper accessed object using techniques such as code static analysis and will complete the member access expression.

\section{Methodology}


We explain the details of our code generation method in this section. In section \ref{subsection:methodology:task_augmentation}, we present our method of applying a variable semantic table for task augmentation. In section \ref{subsection:methodology:subtoken_tranx}, we  describe the architecture of our Subtoken-TranX model. In section \ref{subsection:methodology:js_support}, we describe the details of how we make Subtoken-TranX support JavaScript language, which also applies in the original TranX model.

\subsection{Task Augmentation}
\label{subsection:methodology:task_augmentation}

As we face insufficient training data, we want to leverage external domain knowledge to assist with code generation. We found that variable names occupy a large part of the front-end JavaScript code we study. Thus, it benefits a lot if we make the model learn more about the correct use of variable names from the domain knowledge. Following this idea, we think it is helpful if we can obtain a paired dataset of variable names and their semantic meaning and incorporate this data into the code generation model. We extract a variable semantic table containing the name and semantic description of variables commonly used in front-end development. We collect these variable--description paired data from various sources, including the following:
\begin{itemize}
    \item Requirement documents. The requirement documents may contain some variable name conventions for variables used across modules in the project. We can gather these variable name conventions and add them to the variable semantic table.
    \item Codebase. In the development of large-scale projects, developers usually use some protocols for serializing data structures (Protocol Buffer, Thrift, etc.) to describe different data structures. For the fields in the data structures, there are usually comments that describe the meanings of the fields. We can collect these fields and semantic meanings and add them to the variable semantic table.
    \item Database field definitions. Tables in the database may have corresponding documents to describe the semantics of each field in the table. We can also use these fields and semantic meanings to build the variable semantic table.
\end{itemize}

Eventually, we collect a variable semantic table containing 15,525 variables that are commonly used in front-end developments.
Some of the examples are shown in table \ref{table:semantic_table_example}.

\begin{CJK*}{UTF8}{gbsn}
\begin{table}
    \centering
    \begin{tabular}{ll}
        \toprule
        Variable Name        & Variable Semantic                    \\
        \midrule
        shopLogo             & 店铺标志(shop logo)                  \\
        beforePromotionPrice & 促销前价格(price before promotion)   \\
        storeCityName        & 门店所属城市(city the store located) \\
        roomStatus           & 直播状态(live room status)           \\
        picUrl               & 图片链接(link of the picture)        \\
        $\cdots$             & $\cdots$                             \\
        \bottomrule
    \end{tabular}
    \caption{Examples of the variable semantic table}
    \label{table:semantic_table_example}
    \st
\end{table}
\end{CJK*}

There are many ways to incorporate the variable semantic table into the code generation model. For example, we can use the variable semantic table to pre-train the model and then use the code generation data to fine-tune the model. In our attempts, we use a task augmentation method to leverage the variable semantic table for code generation.

Task augmentation is the way of training a neural network with multiple tasks. In that way, the model can learn more knowledge from the auxiliary tasks and improve the main task's performance. In our practice, we take the variable semantic table as domain knowledge and design auxiliary tasks to train the model with the main code generation task. This task augmentation method has two main benefits. First, the model can learn more about the relations between variable semantics and variable names from the auxiliary task and use it in code generation. Thus the model may make better predictions of variable names during code generation and improve the overall performance. Second, the addition of auxiliary tasks enlarges the training data size and reduces the overfitting of the model.

Many different auxiliary tasks can leverage the variable semantic table as training data. The most straightforward task is to predict variable names from variable semantic meanings. This task requires the model to understand the variable semantic meanings and predict the correct variable name. However, the output of this auxiliary task is not a complete code, and naturally, we can not convert it into a legal abstract syntax tree. So we cannot apply this auxiliary task to tree-based code generation methods such as TranX. We can only use this auxiliary task on token sequence-based methods.

\begin{CJK*}{UTF8}{gbsn}
In order to apply the task augmentation in AST-based code generation methods, we propose another auxiliary task. We found that for a string template expression that has no other string literals to join with, the code is just a variable name plus syntax symbols such as braces and semicolons. We can transform the variable semantic table into these string template expressions. Taking the ``picUrl{--}图片链接(link of the picture)'' as an example, we can rewrite the input as ``展示图片链接(show link of the picture)'' and the output as ``\texttt{\{ picUrl; \}}''. In that way, we write the output to a legal code and can be parsed to an abstract syntax tree so that we can apply the auxiliary task to AST-based code generation methods. Through that rewrite, we have also largely narrowed the gap between the main task and auxiliary task, enabling task augmentation methods to be more effective. We adopt this auxiliary task in our code generation model. We will compare this auxiliary task with the variable names predicting task for token sequence-based code generation on the Transformer model. We'll also compare the task augmentation method with the pre-train method.
\end{CJK*}

\subsection{Subtoken-TranX Model}
\label{subsection:methodology:subtoken_tranx}


The original TranX model generates code at the token level. In that way, an identifier corresponds to a token in the vocabulary. This setting is not friendly to code generation under small-scale training data. The token-level vocabulary is usually large, so the model needs to maintain a large embedding matrix, which makes the model easy to overfit. And the identifiers in code may appear only a few times in the training set, so it is hard for the model to learn a good word embedding representation. Besides, the model may fail to capture the relationship of tokens containing the same subtoken from limited training data. A better way is to generate code at the subtoken level.

To generate code at the subtoken level, we subtokenize all the identifiers according to case boundaries for identifiers in the camel case and underscores for identifiers in the snake case \cite{binkley2009camelcase} for all the code. For example, the identifier \texttt{liveTimeDesc} will be split to \texttt{live}, \texttt{\#\#Time}, and \texttt{\#\#Desc}. The prefix \texttt{\#\#} indicates that this subtoken and the previous one are split from the same token, and we need to join them together when converting subtokens back to tokens. We choose not to use statistical-based subtokenization methods such as BPE \cite{sennrich2016neural} because we think the distribution of small-scale corpus is not statistically significant. Thus, the subtokens learned with BEP from the small-scale corpus may not truly reflect the actual distribution of subtokens and may not achieve the best performance.

To perform the subtoken-level code generation, we build our Subtoken-TranX model based on the original TranX model. There are three stages to transition from natural language description to code in Subtoken-TranX. First, the encoder-decoder neural network in TranX takes the natural language description as input and then generates a sequence of actions to construct the AST. In this stage, the model will check the syntax of the target programming language and make sure the generated actions can be used to build a valid AST. Second, we construct the AST with the generated action sequence in the first stage. Third, we convert the AST to target code. The AST-constructing actions defined in the original TranX model have three types:
\begin{itemize}
    \item \textsc{ApplyConstr}$[c]$. This action applys a construction rule $c$ on current field of the AST under construction.
    \item \textsc{Reduce}. This action marks the end of generation of the current field with optional or multiple cardinalities.
    \item \textsc{GenToken}$[v]$. This action generates a terminal token as a leaf node to the current field.
\end{itemize}

\begin{figure*}
    \centering
    \begin{subfigure}[b]{0.49\linewidth}
        \centering
        \vspace{0.1cm}
        \resizebox{\textwidth}{!}{Code: \texttt{\{contentType === 'live' ? liveTimeDesc : marketingTimeDesc\}}}

        \vspace{0.2cm}
        \includegraphics[width=0.95\textwidth]{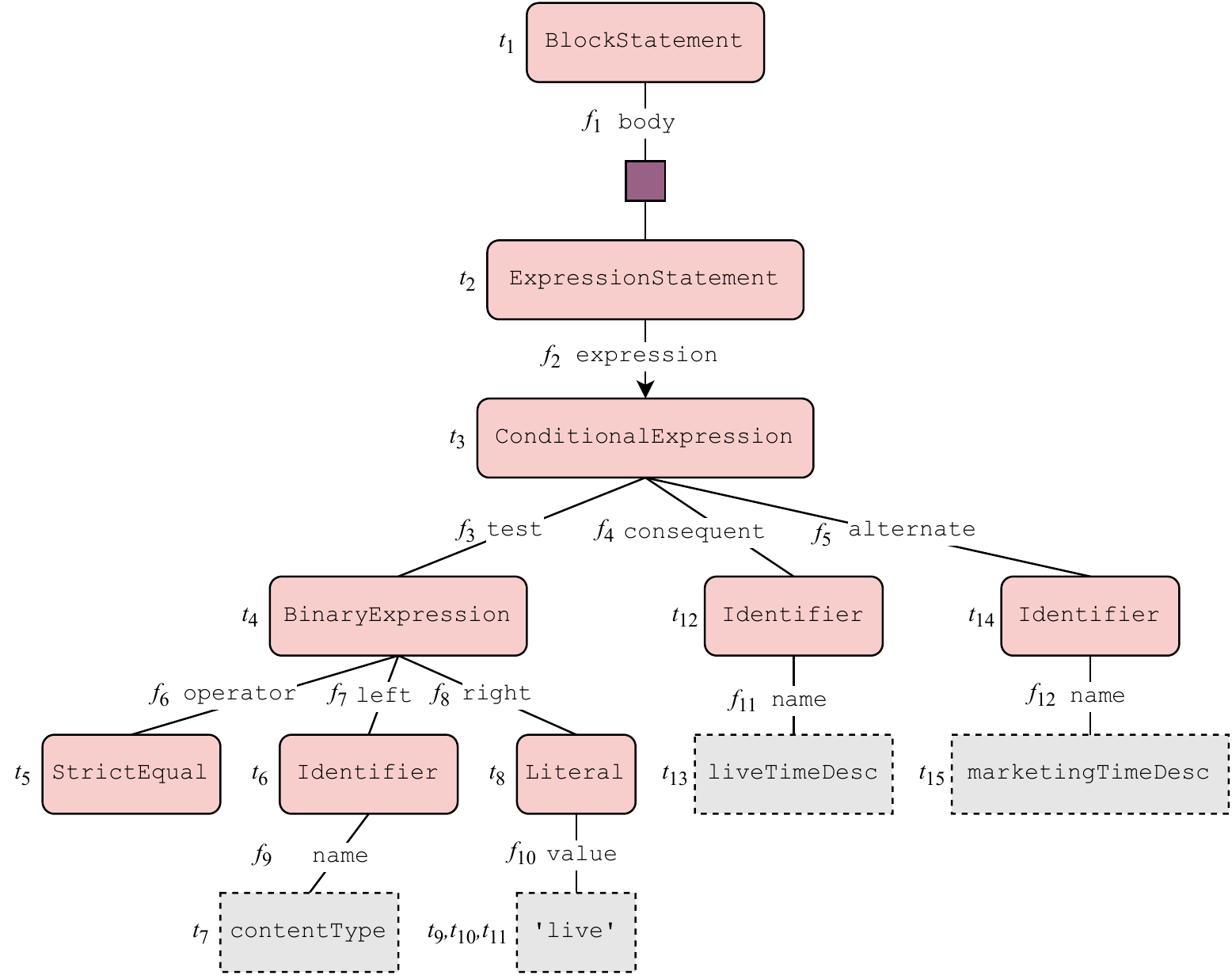}
        \label{figure:subtoken_tranx_example_ast}
    \end{subfigure}
    \hspace{\fill}
    \begin{subfigure}[b]{0.49\textwidth}
        \centering
        \resizebox{\textwidth}{!}{
            \begin{tabular}{lll}
                \toprule
                $t$        & $n_{f_t}$     & Action                                                                      \\
                \midrule
                $t_1$      & \texttt{root} & \texttt{BlockStatement(stmt* body)}                                         \\
                $t_2$      & $f_1$         & \texttt{ExpressionStatement(expr expression)}                               \\
                $t_3$      & $f_2$         & \texttt{ConditionalExpression(expr test, expr alternate, expr consequent)}  \\
                $t_4$      & $f_3$         & \texttt{BinaryExpression(binary\_operator operator, expr left, expr right)} \\
                $t_5$      & $f_6$         & \texttt{StrictEqual()}                                                      \\
                $t_6$      & $f_7$         & \texttt{Identifier(name)}                                                   \\
                $t_{7,1}$  & $f_9$         & \textsc{GenSubtoken}[\texttt{content}]                                      \\
                $t_{7,2}$  & $f_9$         & \textsc{GenSubtoken}[\texttt{\#\#Type}]                                     \\
                $t_{7,3}$  & $f_9$         & \textsc{GenSubtoken}[\texttt{<EOT>}]                                        \\
                $t_8$      & $f_8$         & \texttt{Literal(literal? value)}                                            \\
                $t_9$      & $f_{10}$      & \textsc{GenSubtoken}[\texttt{<SOS>}]                                        \\
                $t_{10}$   & $f_{10}$      & \textsc{GenSubtoken}[\texttt{live}]                                         \\
                $t_{11}$   & $f_{10}$      & \textsc{GenSubtoken}[\texttt{<EOS>}]                                        \\
                $t_{12}$   & $f_4$         & \texttt{Identifier(identifier name)}                                        \\
                $t_{13,1}$ & $f_{11}$      & \textsc{GenSubtoken}[\texttt{live}]                                         \\
                $t_{13,2}$ & $f_{11}$      & \textsc{GenSubtoken}[\texttt{\#\#Time}]                                     \\
                $t_{13,3}$ & $f_{11}$      & \textsc{GenSubtoken}[\texttt{\#\#Desc}]                                     \\
                $t_{13,4}$ & $f_{11}$      & \textsc{GenSubtoken}[\texttt{<EOT>}]                                        \\
                $t_{14}$   & $f_{5}$       & \texttt{Identifier(identifier name)}                                        \\
                $t_{15,1}$ & $f_{12}$      & \textsc{GenSubtoken}[\texttt{marketing}]                                    \\
                $t_{15,2}$ & $f_{12}$      & \textsc{GenSubtoken}[\texttt{\#\#Time}]                                     \\
                $t_{15,3}$ & $f_{12}$      & \textsc{GenSubtoken}[\texttt{\#\#Desc}]                                     \\
                $t_{15,4}$ & $f_{12}$      & \textsc{GenSubtoken}[\texttt{<EOT>}]                                        \\
                $t_{16}$   & $f_1$         & \textsc{Reduce} (close the frontier field $f_1$)                            \\
                \bottomrule
            \end{tabular}
        }
        \label{figure:subtoken_tranx_example_actions}
    \end{subfigure}
    \caption{An example of JavaScript AST and the action sequence to generate the AST. In this example, the generation of token \texttt{liveTimeDesc} of field $f_{11}$ and \texttt{marketingTimeDesc} of field $f_{12}$ is split into several actions with each action generates a subtoken.}
    \label{figure:subtoken_tranx_example}
\end{figure*}

In the original TranX model, the generation of a terminal token corresponds to one and only one \textsc{GenToken} action, so the original TranX model only supports token-level code generation but not subtoken-level code generation. To enable the model to support subtoken-level generation, we need to do some modifications to the \textsc{GenToken} action. We replace the \textsc{GenToken} action with \textsc{GenSubtoken} in the Subtoken-TranX model. We stipulate that at each position of AST that needs to generate a terminal symbol, the model can generate multiple \textsc{GenSubtoken} actions, of which each \textsc{GenSubtoken} action generates a subtoken. We use a special token \texttt{<EOT>} to indicate that all the subtokens of the token in the current field have been completely generated. The subtokens generated by these consecutive \textsc{GenSubtoken} actions will be joined together to form the token generated at the current position and inserted into the AST under construction. The model will then proceed to the generation of the next field. With this modification, the Subtoken-TranX model can support subtoken-level code generation. Figure \ref{figure:subtoken_tranx_example} show an example of JavaScript AST and the sequence of actions used to construct the AST. In this example, the identifier \texttt{contentType} contains two subtokens which correspond to the \textsc{GenSubtoken} actions in time step $t_{7,1}$ and $t_{7,2}$. In time step $t_{7,3}$, the model generates the \textsc{<EOT>}, then it merges all the consecutive subtokens generated in this field to the token \texttt{contentType} and inserts it to the AST under construction. The generation of \texttt{liveTimeDesc} and \texttt{marketingTimeDesc} is also similar.

Noticing that this modification is target programming language agnostic, the Subtoken-TranX model can be used to generate code in other programming languages besides JavaScript. We can replace the original TranX model with the Subtoken-TranX model whenever subtoken-level generation is preferred.

The architecture of neural networks in Subtoken-TranX follows the original TranX in \cite{yin2018tranx}. We compute the probabilities of generating an action sequence $\mathbf{z}$ as
\begin{equation}
    p(\mathbf{z}|\mathbf{x}) = \prod_t p(a_t|a_{<t}, \mathbf{x}).
\end{equation}

The encoder is a bidirectional LSTM \cite{hochreiter1997long} which encodes the input utterance $\{x_i\}_{i=1}^n$ into representations $\{\mathbf{h}_i\}_{i=1}^n$. The decoder is also an LSTM network which computes the hidden state $\mathbf{s}_t$ at each time step as
\begin{equation}
    \mathbf{s}_t=f_{\mathrm{LSTM}}\left(\left[\mathbf{a}_{t-1}; \tilde{\mathbf{s}}_{t-1}; \mathbf{p}_t\right], \mathbf{s}_{t-1}\right).
\end{equation}
The inputs of decoder LSTM are three vectors. $\mathbf{a}_{t-1}$ is the embedding of last action. $\tilde{\mathbf{s}}_{t-1}$ is the attention result computed as
\begin{equation}
    \tilde{s}_t=\tanh\left(W_c\left[c_t;s_t\right]\right),
\end{equation}
where $c_t$ is the context vector computed in attention \cite{luong2015effective} with encoded results ${\mathbf{h}_i}_{i=1}^n$. $\mathbf{p}_t$ is parent feeding information which is the concatenation of the embedding of the frontier field $\mathbf{n}_{f_t}$ and $\mathbf{}{s}_{p_t}$, the decoder's state at which the constructor of $n_{f_f}$ is generated by the \textsc{ApplyConstr} action.

We compute the probability of action \textsc{ApplyConstr}$[c]$ as
\begin{equation}
    p(a_t=\textsc{ApplyConstr}[c] \left|\right. a_{<t}, \mathbf{x})=\mathrm{softmax}\left(\mathbf{a}_c^\top \mathbf{W} \tilde{\mathbf{s}}_t\right).
\end{equation}
The probability of \textsc{GenSubtoken} is a hybrid probability of generation and copy from input, which is formulated as
\begin{equation}
    \begin{aligned}
        p(a_t=\textsc{GenSubtoken}[v] \left|\right. a_{<t}, \mathbf{x})=p(\mathrm{gen} | a_t, \mathbf{x})p(v|\mathrm{gen}, a_t, \mathbf{x}) \\ + p(\mathrm{copy} | a_t, \mathbf{x})p(v|\mathrm{copy}, a_t, \mathbf{x}),
    \end{aligned}
\end{equation}
where
\begin{equation}
    p(v|\mathrm{gen}, a_t, \mathbf{x})=\mathrm{softmax}\left(\mathbf{a}_c^\top \mathbf{W} \tilde{\mathbf{s}}_t\right),
\end{equation}
\begin{equation}
    p(v|\mathrm{copy}, a_t, \mathbf{x})=\mathrm{softmax}\left(\mathbf{h}_t^\top \mathbf{W} \tilde{\mathbf{s}}_t\right),
\end{equation}
and
\begin{equation}
    p(\mathrm{gen}|\cdot), p(\mathrm{copy}|\cdot)=\mathrm{softmax}\left(\mathbf{W}\tilde{\mathbf{s}}_t\right).
\end{equation}
We add beam search on the model so that the model will predict top-$k$ results for every input utterance.

\subsection{JavaScript Language Support}
\label{subsection:methodology:js_support}

The TranX model (and our Subtoken-TranX model) uses ASDL \cite{wang1997zephyr} to describe the syntax of programming languages. Most of the previous works that use the TranX model for general-purpose programming language code generation \cite{yin2018tranx,jiang2021exploring,xie2021improving} use Python as the target language. The Python language officially uses ASDL to describe the syntax rules. So it is convenient to utilize the TranX model for Python code generation. For the Subtoken-TranX model to support the generation of the JavaScript code, we need to construct the ASDL for the syntax rules of the JavaScript language.

ASDL is a relatively simple abstract syntax description language. It only supports using exactly one type to constrain a field and does not support using multiple types to constrain a field. It also does not support using subtype to constrain a field. For programming languages with complex syntax rules such as JavaScript, we can only support parts of the syntax rules of JavaScript and make particular adaptations for some of the syntaxes.

Table \ref{code:js_incompat} show 2 examples of JavaScript's abstract syntax that are incompatible with ASDL. In the \texttt{CallExpression} of JavaScript, the type of field \texttt{callee} can be either \texttt{Expression} or \texttt{Import}. Given that the ASDL only supports using exactly one type to constrain a field, we just keep the \texttt{Expression} type for the field and remove support for \texttt{Import} type for mostly the \texttt{callee} fields are of type \texttt{Expression}. We do this kind of choice for all the syntaxes with multiple types for a field and keep only the most frequently used type. In \texttt{BreakStatement}, the \texttt{label} field is specified as \texttt{Identifier}, which is not a type but a subtype of \texttt{Statement} (or \texttt{stmt} in our ASDL since we use lower case abbreviations to represent types). Now that ASDL does not support using subtype to constrain a field, we directly use its parent type \texttt{stmt} as the constrain instead of \texttt{Identifier} to comply with the rules of ASDL. We relax the type constraints in this way for all syntaxes that has the subtype issue. This treatment brings up another problem that the generated AST containing these syntaxes may comply with ASDL syntax constraints but is not a legal JavaScript AST. If we encounter this situation during code generation, we simply discard this result and adopt other results predicted by the model.

\begin{table}
    \centering
    \begin{tabular}{l}
        \toprule
        Syntax of \texttt{CallExpression}:                           \\
        \texttt{\makecell[l]{
        interface CallExpression \{                                  \\
        \hspace{2em}type: 'CallExpression';                          \\
        \hspace{2em}callee: Expression | Import;                     \\
        \hspace{2em}arguments: ArgumentListElement[];                \\
        \}                                                           \\
        }}                                                           \\
        \\
        ASDL of \texttt{CallExpression}:                             \\
        \texttt{expr = CallExpression(expr callee, expr* arguments)} \\
        \midrule
        Syntax of \texttt{BreakStatement}:                           \\
        \texttt{\makecell[l]{                                        \\
        interface BreakStatement \{                                  \\
        \hspace{2em}type: 'BreakStatement';                          \\
        \hspace{2em}label: Identifier | null;                        \\
        \}                                                           \\
        }}                                                           \\
        \\
        ASDL of \texttt{BreakStatement}:                             \\
        \texttt{stmt = BreakStatement(expr? label)}                  \\
        \bottomrule
    \end{tabular}
    \caption{Examples of ASDL-incompatible JavaScript abstract syntax}
    \label{code:js_incompat}
    \st
\end{table}

We write the ASDL of JavaScript regarding the JavaScript abstract syntax in the form of Mozilla Parser API provided in the documentation of Esprima parser. The ASDL of JavaScript we write can cover the vast majority of the front-end JavaScript code in our study scenario. 

In addition to the ASDL of JavaScript, we write two functions to do the conversion between ASDL AST and JavaScript AST. The functions traverse the input AST and build the output AST according to the information of AST nodes. We use Esprima and Escodegen to do the conversion between JavaScript AST and JavaScript code. So far, we have added JavaScript language support to the Subtoken-TranX model and can use it to generate JavaScript code from natural language descriptions.

\section{Experimental Settings}


\subsection{Setup}

\subsubsection{Dataset}

We use the dataset presented above in section \ref{section:dataset} for our experiments. We randomly sample a subset from the training set as our validation set. In the task augmentation, we use the variable semantic table of 15,525 variables that we collected from the codebase, database field definitions, and requirement documents.

\subsubsection{Baselines}

Considering that there are two main categories of code generation models, tree-based and sequence-based, we selected a typical model from each category as our baselines. They are:
\begin{itemize}
    \item \textbf{TranX}. The TranX model is an AST-based token-level code generation model and performed well in previous work.
    \item \textbf{Transformer}. The Transformer model is widely-used and achieves good performance in many sequence-to-sequence tasks. We use it for subtoken-level code generation.
\end{itemize}
We compare our proposed Subtoken-TranX model with the two baselines, and we also evaluate our task augmentation method on the Subtoken-TranX and the baseline models.

\subsubsection{Implementation Details}

For the Subtoken-TranX and baseline TranX model, we use the tools Esprima and Escodegen for the conversion between JavaScript code and AST. We also use Esprima to tokenize the JavaScript code for the baseline Transformer. We tokenize the natural language description with NLTK and Jieba. We subtokenize the identifier according to the case boundaries for identifiers in the camel case and the underscores for those in the snake case. The Subtoken-TranX model is built based on the open-source code of TranX \footnote{\url{https://github.com/pcyin/tranX}}, and the Transformer model is built with the official implementation of the Transformer model in PyTorch.

For the Subtoken-TranX and TranX models, we set the hidden size of the encoder and decoder LSTM to 256. The embedding size of subtokens and actions is 128. During the training process, we use a batch size of 32 and train the model for 300 epochs. We use Adam \cite{kingma2015adam} optimizer and set the initial learning rate to $1\times 10^{-3}$.

For the Transformer model, we use the encoder and decoder with 4 layers. We set the model size to 128, and the size of the feed-forward network to 512. Each multi-head attention in the Transformer model has 4 heads. During training, we use a batch size of 32 and train the model for 300 epochs. We use the AdamW \cite{loshchilov2018decoupled} optimizer to optimize the model parameters. We use the learning rate warming up and decay during the training process and set the max learning rate to $1\times 10^{-4}$.

We add beam search on both Subtoken-TranX and Transformer and set the beam width to 5 for both models.

\subsection{Evaluation Metrics}
We use the following evaluation metrics for the code generation:
\begin{itemize}
    \item \textbf{Exact Match Accuracy}. The exact match accuracy regards a predicted result as true when the result is exactly the same as the reference code. This is a very strict metric. For codes that are functionally identical but literally different, the exact match accuracy regards the prediction as false. 
    \item \textbf{BLEU} \cite{papineni2002bleu}. The BLEU score checks the precision of n-grams in the predicted result and computes a composite score. We use the \texttt{corpus\_bleu} in NLTK to compute the BLEU score of our result.
    \item \textbf{Edit Similarity}. The edit similarity computes the similarity of two strings with Levenshtein edit distance.
    When used in code generation, this metric indicates how much modification the user needs to change the generated code to the correct reference code.
\end{itemize}

\section{Experimental Results}

\subsection{Main Results}

To verify the effectiveness of our task augmentation method that leverages external domain knowledge for code generation and our Subtoken-TranX model, we train the models with and without task augmentation and compare the results of different models on code generation. Table \ref{table:main_result} show the main results of code generation. The results with ``+TA'' are the results when training the models with task augmentation. The results show that the Subtoken-TranX outperforms the original TranX and Transformer model with or without the task augmentation applied. When applied with task augmentation, both the Transformer and the Subtoken-TranX model that generate code at the subtoken level have significantly improved their performance. The top-1 accuracy and top-5 accuracy have improved by more than 10\%. The results demonstrate the effectiveness of our task augmentation method that incorporates external domain knowledge for code generation on subtoken-level generation models. We also notice that the results of TranX decrease after applying the task augmentation. We think this may be because the addition of the semantic mapping table only has a limited effect on the token-level generation model. On the contrary, the increase in the vocabulary size will increase the number of models and make it easier to overfit so that the model performance will decrease.

\begin{table}
    \centering
    \begin{tabular}{lllll}
        \toprule
        \textbf{Method}   & \textbf{Acc-1} & \textbf{Acc-5} & \textbf{BLEU}  & \textbf{EditSim} \\
        \midrule
        Transformer       & 16.84          & 26.02          & 67.98          & 77.31            \\
        TranX             & 17.35          & 22.45          & 67.56          & 76.20            \\
        Subtoken-TranX    & 20.41          & 29.08          & 66.77          & 77.63            \\
        \midrule
        Transformer+TA    & 29.08          & 39.29          & 69.72          & 82.92            \\
        TranX+TA          & 16.33          & 27.55          & 61.66          & 74.77            \\
        Subtoken-TranX+TA & \textbf{33.16} & \textbf{40.31} & \textbf{71.94} & \textbf{85.27}   \\
        \bottomrule
    \end{tabular}
    \caption{Main results of code generation}
    \label{table:main_result}
    \st
\end{table}

\subsection{Results of Variable Usage}

We conjecture that the task augmentation with the variable semantic table enables the model to learn more about the relationship between variable semantics and variable names from auxiliary tasks, thereby improving the accuracy of variable usage during code generation. To verify our conjecture, we verify the results of variable usage in code generation. We extract all variable names in the predicted results and the reference code on the test set and calculate the precision, recall, and F1 of the variable usage. Table \ref{table:result_identifier} show the results of variable usage in code generation. We can see that the metrics of variable usage have great improvement on all the models after applying the task augmentation. The results validate our conjecture that the addition of auxiliary tasks can improve the accuracy of variable usage in code generation. Among all the models, the Subtoken-TranX model achieves the best precision, recall, and F1 for variable usage.

\begin{table}
    \centering
    \begin{tabular}{llll}
        \toprule
        \textbf{Method}   & \textbf{Precision} & \textbf{Recall} & \textbf{F1}    \\
        \midrule
        Transformer       & 27.66              & 27.27           & 27.46          \\
        TranX             & 26.67              & 27.97           & 27.30          \\
        Subtoken-TranX    & 30.57              & 33.57           & 32.00          \\
        \midrule
        Transformer+TA    & 41.99              & 41.26           & 41.62          \\
        TranX+TA             & 36.12              & 37.76           & 36.92     
        \\
        Subtoken-TranX+TA & \textbf{50.00}     & \textbf{48.25}  & \textbf{49.11} \\
        \bottomrule
    \end{tabular}
    \caption{Results of variable usage in code generation.}
    \label{table:result_identifier}
    \st
\end{table}

\subsection{Results of Different Categories}

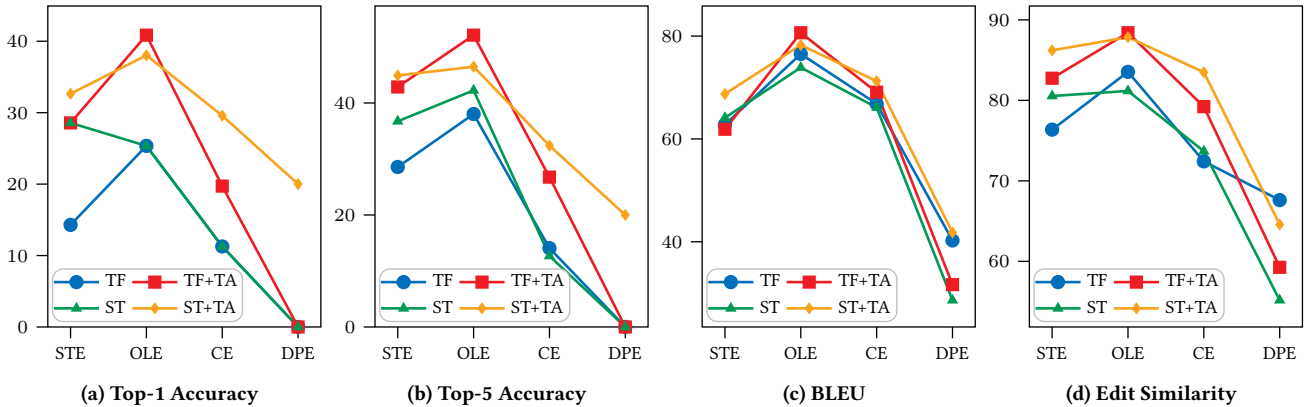
\begin{figure*}
    \begin{subfigure}[b]{0.24\textwidth}
        \begin{tikzpicture}
            \tikzstyle{every node}=[font=\footnotesize]
            \begin{axis}
                [
                    scale only axis,
                    width=0.85\textwidth,
                    height=\textwidth,
                    title style={font=\normalsize\bfseries},
                    axis line style = {line width=0.5pt},
                    tick align=outside,
                    xtick pos=left,
                    ytick pos=left,
                    tick style = {color=black, line width=0.5pt},
                    xtick=data,
                    enlarge x limits = true,
                    ymin=0,
                    legend columns=2,
                    legend cell align = left,
                    legend style={at={(0.015,0.015)},anchor=south west, draw=lightgray, line width=0.5pt, rounded corners, inner sep=0pt, font=\scriptsize},
                    cycle list name=my black white,
                    symbolic x coords = {STE, OLE, CE, DPE},
                ]
                \addplot coordinates {(STE, 14.29) (OLE, 25.35) (CE, 11.27) (DPE, 0.00)};
                \addplot coordinates {(STE, 28.57) (OLE, 40.85) (CE, 19.72) (DPE, 0.00)};
                \addplot coordinates {(STE, 28.57) (OLE, 25.35) (CE, 11.27) (DPE, 0.00)};
                \addplot coordinates {(STE, 32.65) (OLE, 38.03) (CE, 29.58) (DPE, 20.00)};
                \legend{TF, TF+TA, ST, ST+TA}
            \end{axis}
            \pgfresetboundingbox
            \path
            (current axis.south west) -- ++(-0.2in,-0.2in)
            rectangle (current axis.north east) -- ++(0in,0.1in);
        \end{tikzpicture}
        \caption{Top-1 Accuracy}
        \label{img:category_result_top1}
    \end{subfigure}
    \begin{subfigure}[b]{0.24\textwidth}
        \begin{tikzpicture}
            \tikzstyle{every node}=[font=\footnotesize]
            \begin{axis}
                [
                    scale only axis,
                    width=0.85\textwidth,
                    height=\textwidth,
                    title style={font=\normalsize\bfseries},
                    axis line style = {line width=0.5pt},
                    tick align=outside,
                    xtick pos=left,
                    ytick pos=left,
                    tick style = {color=black, line width=0.5pt},
                    xtick=data,
                    enlarge x limits = true,
                    ymin=0,
                    legend columns=2,
                    legend cell align = left,
                    legend style={at={(0.015,0.015)},anchor=south west, draw=lightgray, line width=0.5pt, rounded corners, inner sep=0pt, font=\scriptsize},
                    cycle list name=my black white,
                    symbolic x coords = {STE, OLE, CE, DPE},
                ]
                \addplot coordinates {(STE, 28.57) (OLE, 38.03) (CE, 14.08) (DPE, 0.00)};
                \addplot coordinates {(STE, 42.86) (OLE, 52.11) (CE, 26.76) (DPE, 0.00)};
                \addplot coordinates {(STE, 36.73) (OLE, 42.25) (CE, 12.68) (DPE, 0.00)};
                \addplot coordinates {(STE, 44.90) (OLE, 46.48) (CE, 32.39) (DPE, 20.00)};
                \legend{TF, TF+TA, ST, ST+TA}
            \end{axis}
            \pgfresetboundingbox
            \path
            (current axis.south west) -- ++(-0.2in,-0.2in)
            rectangle (current axis.north east) -- ++(0in,0.1in);
        \end{tikzpicture}
        \caption{Top-5 Accuracy}
        \label{img:category_result_top5}
    \end{subfigure}
    \begin{subfigure}[b]{0.24\textwidth}
        \begin{tikzpicture}
            \tikzstyle{every node}=[font=\footnotesize]
            \begin{axis}
                [
                    scale only axis,
                    width=0.85\textwidth,
                    height=\textwidth,
                    title style={font=\normalsize\bfseries},
                    axis line style = {line width=0.5pt},
                    tick align=outside,
                    xtick pos=left,
                    ytick pos=left,
                    tick style = {color=black, line width=0.5pt},
                    xtick=data,
                    enlarge x limits = true,
                    legend columns=2,
                    legend cell align = left,
                    legend style={at={(0.015,0.015)},anchor=south west, draw=lightgray, line width=0.5pt, rounded corners, inner sep=0pt, font=\scriptsize},
                    cycle list name=my black white,
                    symbolic x coords = {STE, OLE, CE, DPE},
                ]
                \addplot coordinates {(STE, 62.77) (OLE, 76.52) (CE, 66.83) (DPE, 40.25)};
                \addplot coordinates {(STE, 61.88) (OLE, 80.65) (CE, 69.08) (DPE, 31.67)};
                \addplot coordinates {(STE, 64.11) (OLE, 73.88) (CE, 66.11) (DPE, 28.63)};
                \addplot coordinates {(STE, 68.75) (OLE, 78.25) (CE, 71.24) (DPE, 41.79)};
                \legend{TF, TF+TA, ST, ST+TA}
            \end{axis}
            \pgfresetboundingbox
            \path
            (current axis.south west) -- ++(-0.2in,-0.2in)
            rectangle (current axis.north east) -- ++(0in,0.1in);
        \end{tikzpicture}
        \caption{BLEU}
        \label{img:category_result_bleu}
    \end{subfigure}
    \begin{subfigure}[b]{0.24\textwidth}
        \begin{tikzpicture}
            \tikzstyle{every node}=[font=\footnotesize]
            \begin{axis}
                [
                    scale only axis,
                    width=0.85\textwidth,
                    height=\textwidth,
                    title style={font=\normalsize\bfseries},
                    axis line style = {line width=0.5pt},
                    tick align=outside,
                    xtick pos=left,
                    ytick pos=left,
                    tick style = {color=black, line width=0.5pt},
                    xtick=data,
                    enlarge x limits = true,
                    legend columns=2,
                    legend cell align = left,
                    legend style={at={(0.015,0.015)},anchor=south west, draw=lightgray, line width=0.5pt, rounded corners, inner sep=0pt, font=\scriptsize},
                    cycle list name=my black white,
                    symbolic x coords = {STE, OLE, CE, DPE},
                ]
                \addplot coordinates {(STE, 76.36) (OLE, 83.54) (CE, 72.42) (DPE, 67.61)};
                \addplot coordinates {(STE, 82.75) (OLE, 88.40) (CE, 79.22) (DPE, 59.26)};
                \addplot coordinates {(STE, 80.54) (OLE, 81.17) (CE, 73.68) (DPE, 55.17)};
                \addplot coordinates {(STE, 86.21) (OLE, 87.85) (CE, 83.49) (DPE, 64.58)};
                \legend{TF, TF+TA, ST, ST+TA}
            \end{axis}
            \pgfresetboundingbox
            \path
            (current axis.south west) -- ++(-0.2in,-0.2in)
            rectangle (current axis.north east) -- ++(0in,0.1in);
        \end{tikzpicture}
        \caption{Edit Similarity}
        \label{img:category_result_edit_sim}
    \end{subfigure}

    \caption{Metrics of different categories on test data}
    \label{img:category_result}
\end{figure*}

We calculated the metrics of methods on different categories of test data. As mentioned above, we divide the data into four categories: string template expression, OR logic expression, condition expression, and data processing expression. Figure \ref{img:category_result} shows the result of the four categories. We only show the results of Subtoken-TranX and Transformer, as the TranX and Subtoken-TranX models share a similar architecture, and the latter one performs better.

\newcommand\colorone{NavyBlue}
\newcommand\colortwo{Red}
\newcommand\colorthree{ForestGreen}
\newcommand\colorfour{YellowOrange}
\pgfplotscreateplotcyclelist{my black white}{
    solid, every mark/.append style={fill=\colorone, color=\colorone}, mark=*, draw=\colorone, line width=1pt, mark size=2.2pt \\
    solid, every mark/.append style={fill=\colortwo, color=\colortwo}, mark=square*, draw=\colortwo, line width=1pt, mark size=2pt \\
    solid, every mark/.append style={fill=\colorthree, color=\colorthree}, mark=triangle*, draw=\colorthree, line width=1pt, mark size=1.5pt \\
    solid, every mark/.append style={fill=\colorfour, color=\colorfour}, mark=diamond*, draw=\colorfour, line width=1pt, mark size=1.5pt \\
}

Due to the category of data processing expression only having 5 examples in the test set, results on this category of data are not statistically significant, so we focus more on the results of the other three categories. Overall, the performance of the string template expressions and OR logic expressions is better than that of the condition expressions and data processing expressions.
This is because the difficulty of generating condition expressions and data processing expressions condition expressions is greater than that of the other two categories. The condition expressions and data processing expressions may involve more complex logic, and the average number of tokens is significantly longer. 
Besides, there are only a few data processing expressions for training, so the generation of this category is even harder.

We can see from the results that after applying the task augmentation method, the performance of models have different magnitudes of improvement on various categories of data. The results also show that the Subtoken-TranX model outperforms the Transformer model on three categories of data except OR logic expressions. This inconsistency may be because the natural language descriptions of condition expressions have more complex semantics, and the stronger representation ability of the Transformer model can better capture the semantics in natural language descriptions, thereby generating better results.

For the string template expressions, the top-5 accuracy of Subtoken+TA is 44.90\%, and the editing similarity reaches 86.21\%. For the OR logic expressions, the top-5 accuracy of Transformer+TA is 52.11\%, and the editing similarity reached 88.40\%. The model performance on these two code categories has met the requirements for application on actual industrial systems. Our research achievements have been adopted to Alibaba's \emph{BizCook} system.

\subsection{Results of Different Usages of Variable Semantic Table}

In section \ref{subsection:methodology:task_augmentation}, we mentioned that there are many different ways to leverage the variable semantic table for code generation. A straightforward approach in the task augmentation method is to use it for an auxiliary task that predicts variable names from variable semantic meanings. Even though we cannot use this auxiliary task on AST-based code generation methods such as Subtoken-TranX, we can use it on token sequence-based methods such as Transformer. We compare the code generation results using different auxiliary tasks for task augmentation. Besides the task augmentation method, we can also use the variable semantic table for pre-training and fine-tune the model with the code generation dataset.

Table \ref{table:ablation_study_main_result} show the code generation results of incorporating the variable semantic table into the model in different ways. The line of Transfomer is the result of Transformer model training only with code generation dataset. The ``+PT'' is the result of \textbf{\underline{p}}re-\textbf{\underline{t}}raining model with variable semantic table and fine-tuning with code generation dataset. The ``+VP'' is the result of training model with the auxiliary task of \textbf{\underline{v}}ariable name \textbf{\underline{p}}rediction from variable semantics. The ``+CG`` is the result of training model with the auxiliary task after rewriting the variable semantic table into \textbf{\underline{c}}ode \textbf{\underline{g}}eneration paired data. The results show that all the methods of using the variable semantic table are effective in improving code generation performance. In line with our expectation, the task augmentation method with the code generation auxiliary task (+CG) achieves the best performance. It outperforms the variable prediction task augmentation (+VP) because the auxiliary task and the main task have a smaller gap. Similarly, in the pre-train method (+PT), since the model only needs to learn one task in the fine-tuning stage, this method also performs better than variable prediction task augmentation (+VP).

\begin{table}
    \centering
    \begin{tabular}{lllll}
        \toprule
        \textbf{Method} & \textbf{Acc-1} & \textbf{Acc-5} & \textbf{BLEU} & \textbf{EditSim} \\
        \midrule
        Transformer     & 16.84          & 26.02          & 67.98         & 77.31            \\
        Transformer+PT  & 26.53          & 38.27          & 69.91         & 82.47            \\
        Transformer+VP  & 25.00          & 37.24          & 68.82         & 82.38            \\
        Transformer+CG  & 29.08          & 39.29          & 69.72         & 82.92            \\
        \bottomrule
    \end{tabular}
    \caption{Results of different uses of variable semantic table}
    \label{table:ablation_study_main_result}
    \st
\end{table}

\subsection{Case Study}

We conducted a sample analysis to compare the generated results of different methods visually. We picked two individual examples from the test dataset to show the generation results of models.

Table \ref{table:case_study_1} shows the first example. We can see from the results that both the Subtoken-TranX and Transformer models failed to predict the variable \texttt{trainHeadTitle} without applying the task augmentation method. The models may capture some related information from the natural language and generate corresponding subtokens such as \texttt{head} and \texttt{title}. But finally, they fail to combine all the subtokens in the correct way to generate the correct variable name. We examine the variable semantic table we extracted and found the variable \texttt{trainHeadTitle} exists in the table. After applying the task augmentation, the model can learn the correct variable name from the auxiliary tasks. Thus, both Subtoken-TranX and Transformer generate the correct result in the top-1 prediction.
\begin{CJK*}{UTF8}{gbsn}
\begin{table*}
    \centering
    \begin{tabular}{ll}
        \toprule
        Description              & 动态展示火车头标题，兜底显示‘春运火车票’                                                    \\
        Desc. Translation        & Display the title of the train head, the default value is ``Spring Festival Train Ticket''. \\
        \midrule
        Reference Code           & \texttt{\{trainHeadTitle || '\vphantom{}春运火车票';\}}                                                \\
        \midrule
        Subtoken-TranX Result    & \texttt{\{downTitle || '\vphantom{}春运火车票';\}}                                                     \\
        Transformer Result       & \texttt{\{headBannerTitle || '\vphantom{}春运火车票';\}}                                               \\
        Subtoken-TranX+TA Result & \texttt{\{trainHeadTitle || '\vphantom{}春运火车票';\}} \ \CheckmarkBold                               \\
        Transformer+TA Result    & \texttt{\{trainHeadTitle || '\vphantom{}春运火车票';\}} \ \CheckmarkBold                               \\
        \bottomrule
    \end{tabular}
    \caption{Case study example 1.}
    \label{table:case_study_1}
\end{table*}
\end{CJK*}

Table \ref{table:case_study_2} shows the second example. We can find that in the Transformer results without task augmentation, most of the results contain the variable \texttt{subTitle} and the number \texttt{16}. We found an example in training data that is very similar to this example. Just replacing the \texttt{title} to \texttt{subTitle} and \texttt{15} to \texttt{16} in this test example, we get the similar code in training data, which is exactly the same with the top-1 prediction of Transformer in this test example. This behavior implies that the Transformer model has large overfitting. The model "remembers" the content of the training data set and directly outputs the same results when encountering similar samples. After applying task augmentation, the top-2 result of Transformer+TA and the top-1 result of Subtoken-TranX+TA gives the correct predictions. This improvement shows that we have alleviated the over-fitting of the model to a certain extent by expanding the size of training data with task augmentation and improving the model's performance.
\begin{CJK*}{UTF8}{gbsn}
\begin{table*}
    \centering
    \resizebox{\textwidth}{!}{
        \begin{tabular}{ll}
            \toprule
            Description                               & 判断是否显示中间，成立则展示标题前15个字，否则展示前10个字或者不展示                                              \\
            Desc. Translation                         & If show the middle , display the first 15 words of the title, otherwise display the first 10 words or not display \\
            \midrule
            Reference Code                            & \texttt{\{isShowMid ? title.substring(0, 15) : title.substring(0, 10) || '';\}}                                   \\
            \midrule
            \multirow{2}{*}{Subtoken-TranX Result}    & R1. \texttt{\{isShowMid ? title.substring(0, null) : title.substring(0, 10) || null;\}}                           \\
                                                      & R2. \texttt{\{isShowMid ? title.substring(0, 15) : title.substring(0, 10) || null;\}}                             \\
            \midrule
            \multirow{2}{*}{Transformer Result}       & R1. \texttt{\{isShowMid ? subTitle.substring(0, 16) : subTitle.substring(0, 10) || '';\}}                         \\
                                                      & R2. \texttt{\{isShowSuccess ? subTitle.substring(0, 16) : subTitle.substring(0, 10) || '';\}}                     \\
            \midrule
            \multirow{2}{*}{Subtoken-TranX+TA Result} & R1. \texttt{\{isShowMid ? title.substring(0, 15) : title.substring(0, 10) || '';\}} \ \CheckmarkBold              \\
                                                      & R2. \texttt{\{isShowMid ? title.substring(0, 15) : title.substring(null, 15) || '';\}}                            \\
            \midrule
            \multirow{2}{*}{Transformer+TA Result}    & R1. \texttt{\{isShowMid ? title.substring(0, 16) : title.substring(0, 10) || ''\}}                                \\
                                                      & R2. \texttt{\{isShowMid ? title.substring(0, 15) : title.substring(0, 10) || ''\}} \ \CheckmarkBold               \\
            \bottomrule
        \end{tabular}
    }
    \caption{Case study example 2.}
    \label{table:case_study_2}
\end{table*}
\end{CJK*}

\section{Related Work}


This paper is related to deep learning-based code generation methods and how to incorporate external knowledge into code generation.

Code generation usually refers to generating code according to a natural language description. Some works also refer to statement-level code completion as code generation \cite{svyatkovskiy2020intellicode}. In natural language processing, the semantic parsing task that converts natural language to a logical form is similar to code generation. The solutions for these two tasks are basically the same. So we make no distinction between related work on these two tasks and call them code generation collectively. Overall, code generation methods with neural networks can be divided into two categories that are token-based methods and tree-based methods. The token-based methods regard the input utterance and output code as token sequences and use seq2seq network \cite{dong2016language,ling2016latent} or Transformer to generate the output sequence from the input sequence. The tree-based methods have various attempts. The seq2tree network \cite{dong2016language} generates a tree-structured representation in top-down and breadth-first order from natural language input. The Abstract Syntax Networks \cite{rabinovich2017abstract} generate an abstract syntax tree in depth-first order by selecting and applying a constructor to the AST under construction. That makes the output tree conform to the programming language syntax. The TranX \cite{yin2018tranx} generates a sequence of actions that construct an AST. It also utilizes the programming language syntax to filter the actions that do not meet the syntax. The model architecture is more simple than ASN and also achieves good performance. Some other improvements to the TranX model have also resulted in improved performance \cite{jiang2021exploring, xie2021improving}.

In addition to innovations in model architectures, researchers also have proposed many ways to incorporate external knowledge into code generation. There are retrieval-based methods that search for a similar code snippet with the input natural language. The code generation model can use the retrieved code to guide the code generation \cite{hayati2018retrieval,zhang2018guiding} or predict the output by editing the retrieved code \cite{hashimoto2018retrieve}. Some works utilize text--code paired corpus mined from Stack Overflow as an external knowledge to pre-train the code generation model \cite{xu2020incorporating}. With the widespread application of pre-training techniques, many works leverage models pre-trained with large-scale corpus data for code generation \cite{clement2020pymt5, lu2021codexglue, ahmad2021unified, wang2021codet5} and achieve remarkable performance.

\section{Conclusion and Future Works}

This paper presents our practice of applying the JavaScript code generation method for front-end development on Alibaba's \emph{BizCook} platform. Facing the problem of insufficient training data, we incorporate external domain knowledge for code generation through task augmentation. We also extend the TranX model and build a Subtoken-TranX model for subtoken-level code generation. We carry out a series of experiments and demonstrate the effectiveness of our methods. The results show that our code generation method has met the application requirements on actual industrial systems in several code categories and has been adopted to Alibaba's \emph{BizCook} system for production.

So far, we can see that the model does not perform as well on condition expressions and data processing expressions as the other two categories. Our current task augmentation method only takes a variable semantic table as domain knowledge. It thus improves the accuracy of variable usage, which is most effective for codes that are sensitive to variable names. For condition expressions, the variable semantic table cannot improve the prediction of conditions, which is an essential part of condition expressions. For data processing expressions, the performance heavily relies on the prediction of data processing API, which is also not presented in the variable semantic table. In future works, we can mine a series of conditions and their natural language descriptions from the codebase and a series of data processing APIs from documents. We can use a similar method as presented in this paper to design auxiliary tasks and leverage these data for task augmentation. In that way, we can improve the code generation performance on condition expressions and data processing expressions. In addition, with the launch of the code generation tool, we can continuously collect the usage history as paired data and use it for model training. The increase in the amount of data will also help improve the performance of the model on various categories of data.


\begin{acks}
This research is supported by Alibaba Group through Alibaba Innovative Research Program and the National Natural Science Foundation of China under Grant No. 62072007, 62192733, 61832009, 62192731, 62192730.
\end{acks}

\newpage

\bibliographystyle{ACM-Reference-Format}
\bibliography{main.bib}

\end{document}